\documentstyle[13lomcon,epsf,epsfig,wrapfig,rotate,graphics,cite]{article}
 \newcommand{\be}{\begin{eqnarray}}
 \newcommand{\ee}{\end{eqnarray}}

\begin{document}

\title{COLLECTIVE EFFECTS IN CENTRAL HEAVY-ION COLLISIONS}

\author{ G.I.Lykasov \footnote{e-mail: lykasov@jinr.ru},
 A.N.Sissakian \footnote{e-mail:  sisakian@jinr.ru},
A.S.Sorin \footnote{e-mail:  sorin@theor.jinr.ru},
V.D.Toneev \footnote{e-mail:  toneev@theor.jinr.ru}}

\address{JINR, Dubna, 141980, Moscow region, Russia}


\maketitle\abstracts {In-medium effects on transverse-mass distributions of quarks and
gluons are considered assuming a possible local equilibrium  for
colorless quark objects (mesons and baryons) created in central
A+A collisions.  It is shown that the average transverse momentum squared
for these partons grows and then saturates
when the initial energy increases. Within the quark-gluon string
model it leads to the colliding energy dependence of hadron transverse mass
spectra which is similar to that observed in heavy ion collisions.
Comparison with other scenarios is given.}

 Experimental detection of the quark-gluon plasma (QGP) phase and 
the mixed phase (MP) in A+A collisions is a nontrivial task because 
of smallness of the space-time volume of the hot and dense  system 
and possible contributions of hadronic processes simulating signals 
of QGP and MP formation. 
Nevertheless, recent experimental study of the transverse-mass spectra 
of kaons from the central $Au+Au$ and $Pb+Pb$ collisions revealed an "anomalous"
dependence on the incident energy. The effective transverse
temperature (the inverse slope-parameter of the transverse mass
distribution at the mid rapidity) increases fast  with
incident energy in the AGS domain~\cite{Ah00}, then saturates at
the SPS energies~\cite{NA49} and  increases again approaching the 
RHIC energy region~\cite{RHIC}. In agreement with
expectations~\cite{Sh80,ShZ80,vH82} this saturation was assumed to
be associated with the deconfinement phase transition and
indication of the MP~\cite{GGB03,Mohanty:2003}. 
The anomalous effective temperature behavior was quite successfully
reproduced within a hydrodynamic model with the equation of state
involving the phase transition~\cite{Braz04}. However, this result
is not very convincing since to fit data, the required
incident-energy dependence of the freeze-out temperature should
closely repeat the shape of the corresponding effective kaon
temperature, and thereby the problem of the observed anomalous
inverse-slope dependence is readdressed to the problem of the
freeze-out temperature.

In this paper, we propose another way to introduce a collectivity
effect in a nuclear system via some  in-medium effect. We consider
the temperature dependence of quark distribution functions inside
a colorless quark-antiquark or quark-diquark system (like meson or
baryon, $h$) created in the central A+A collisions. A contribution of
this effect to transverse momentum spectra of hadron is
estimated and it is shown that  it results in larger values of the
inverse slope parameter  and, therefore, in broadening of the
transverse mass spectra.

Let us assume the local equilibrium  in a fireball of hadrons
whose distribution function can be presented in
the following relativistic invariant form: %
\be f_h^{A}~=~ C_T\left\{1~\pm~exp((p_h\cdot
u-\mu_h)/T)\right\}^{-1}~, \label{def:hdmp} \ee
where  $p_h$ is the four-momentum of the hadron, the four-velocity
of the fireball in the proper system is $u=(1,0,0,0)$, the sign
"$+$" is for fermions and "$-$" is for bosons, $\mu_h$ is the
baryon chemical potential of the hadron $h$, $T$ is the local
temperature, and $C_T$ is the $T$-dependent normalization factor.
The distribution function of constituent quarks inside $h$ which
is in local thermodynamic equilibrium with the surrounding nuclear
matter, $f_q^A(x,{\bf p}_t)$, can be calculated using the procedure
suggested for a free hadron in Ref.\cite{Weiskopf:1971}, see details
in Ref\cite{lsst}, 
 \be f_{q_v}^A(x,p_t)=\int_0^1
dx_1\int_0^1dx_h\int d^2p_{1t}d^2p_{ht} \ {\tilde q}_v(x,{\bf
p}_t) \ {\tilde q}_r(x_1,{\bf p}_{1t})
\\ \nonumber
\times f_h^{A}(x_h,{\bf p}_{ht}) \ \delta(x+x_1-x_h) \
\delta^{(2)}({\bf p}_{1t}+{\bf p}_t-{\bf p}_{ht}) ~,
 \label{def:fqvA}
 \ee
where ${\tilde q}_v(p_z,{\bf p}_t)$ is related to the probability to find
a valence quark with longitudinal momentum $p_z$ and transverse
momentum ${\bf p}_t$ in the hadron, whereas ${\tilde q}_r(p_{1z}, p_{1t})$
is the probability that all the other hadron constituents (one or
two valence quarks plus any number of quark-antiquark $q{\bar q}$
pairs and gluons) carry a total longitudinal momentum $p_{1z}$ and
the total transverse momentum $p_{1t}$; 
$x=2p^*_z/\sqrt{s^\prime}, x_1=2p^*_{1z}/\sqrt{s^\prime},
x_h=2p^*_{hz}/\sqrt{s^\prime}$,  where $p^*_z,p^*_{1z},p^*_{hz}$
are the longitudinal momenta and 
$s^\prime$ is some characteristic energy squared scale.
 
The distribution of the hadron
$h$ in a fireball $f_h^A$ is included in eq.(2), therefore we
integrate over the longitudinal and transverse momenta of $h$. 
Assuming the factorization hypothesis ${\tilde q}_{v,r}(x,{\bf p}_t)=
{\tilde q}_{v,r}(x) \ g_{v,r}({\bf p}_t)$ and the Gaussian form for $g_{v,r}({\bf p}_t)$,
e.g., $g_{v,r}({\bf p}_t)=\sqrt{\gamma_q/\pi}exp(-\gamma_q p_t^2/2)$ we can get the following
expression for $ f_q^{A}(x, {\bf p}_t;T)$ normalized to $1$ \cite{lsst}:
\be f_q^{A}(x, {\bf p}_t;T)=\frac{1}{\pi}\int_0^{1-x} dx_1 \
{\tilde q}_v(x) \ {\tilde q}_r(x_1) \
{\tilde\Gamma}_q(x_1+x)\exp(-{\tilde\Gamma}_q(x_1+x) p_t^2)
\label{def:fqAxapp}
 \ee
where
\be {\tilde\Gamma}_q(x_h)=\frac{\gamma_q(1+\gamma_q{\tilde m}_h(x_h)T/2)} 
{1+\gamma_q {\tilde m}_h(x_h)T}
\label{def:tGm}
\ee
and ${\tilde m}_h(x_h)=\sqrt{m^2_h+x_h s^\prime/4}$.
Then the averaged transverse momentum squared for the quark at $x\simeq 0$ in a 
locally equilibrated hadron is
 \be
<p_{q,t}^2(x\simeq 0)>_{h,appr.}^A&\simeq&\frac{<p_t^2>_q^h+T\sqrt{m_h^2+s^\prime/4}}
{1+T\sqrt{m_h^2+s^\prime/4}/(2<p_t^2>_q^h)}~,
\label{def:aptsqappr}
 \ee
where $<p_t^2>_q^h$ is the transverse momentum squared for the quark in a free hadron.
More accurate calculational results for this quantity of the $u$ quark in a proton 
$<p_{u,t}^2(x\simeq 0)>_p^A$ are presented in Fig.1.
\begin{figure}[t]
\rotatebox{270}%
{\epsfig{file=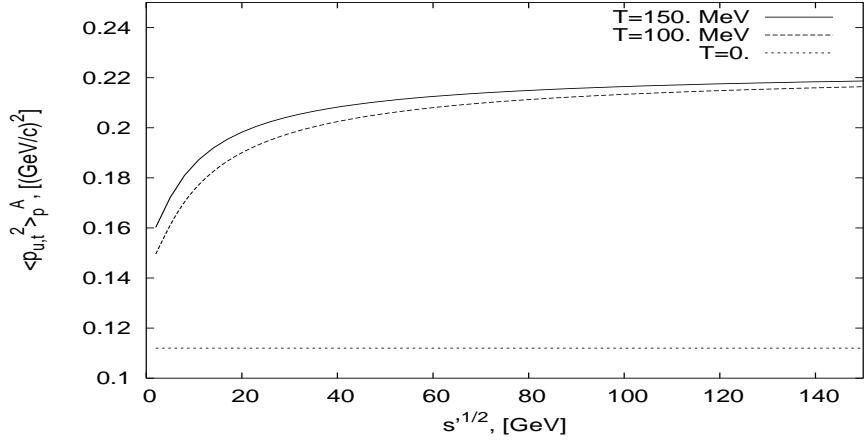,height=12.cm,width=6.cm }}
\caption[Fig.1]{The energy dependence of the average transverse
momentum squared for the $u$-quark in a proton in nuclear matter
at temperature $T$.}
\end{figure}

Let us estimate now the $p_t$ distribution of hadron $h_1$
produced from a collision of two hadrons $h$ inside the fireball.
We shall explore the quark-gluon string model (QGSM) \cite{Kaid1,Kaid2}
or the dual parton model (DPM) \cite{Capella:1994}  
based on the $1/N$ expansion in the QCD \cite{tHooft,Veneziano}. 

To calculate the transverse momentum spectrum of hadron $h_1$ in
the mid rapidity region, one needs to
know the $p_t$-dependence of the fragmentation function
$D_q^{h_1}$. We assume the Gaussian dependence  for $D_q^{h_1}$ like
as for ${\tilde g}_{v,r}(p_t)$. However, the slope of this $p_t$-dependence
$\gamma_c$ can differ from the slope $\gamma_q$ for constituent
quark $p_t$-distribution

 \be
<p_{{h_1}t}^2>_{NN,appr.}^{AA}\simeq
\frac{<p_t^2>_q^N+T\sqrt{m_N^2+s_{hh}/4}}{1+T\sqrt{m_N^2+s_{hh}/4} \ / \
(2<p_t^2>_q^N)}+ \frac{<p_t^2>_q^N}{r}~, \label{def:spthapp}
 \ee
where $s^\prime$ has been associated with the
energy squared $s_{hh}$ of colliding hadrons and $r=\gamma_c/\gamma_q$.
Note, that $\sqrt{s_{hh}}$ is not related directly to the initial energy of 
colliding heavy ions.

We estimated the average value of transverse momentum squared for
$K^+$-mesons produced in the nucleon-nucleon $<p_{K^+,t}^2>_{NN}^{AA}$
and pion+nucleon $<p_{K^+,t}^2>_{\pi N}^{AA}$ interactions 
 in a fireball created in the central $A-A$ collision as a function 
of $\sqrt{s_{hh}}$ at
$T=150 \ MeV$ for two cases when $\gamma_c>>\gamma_q$ 
and $\gamma_c=3\gamma_q$ \cite {LS1}. In Fig.2,  the curves $1$ and $2$ correspond to 
$<p_{K^+,t}^2>_{NN}^{AA}$ and $<p_{K^+,t}^2>_{\pi N}^{AA}$, respectively, when 
$\gamma_c>>\gamma_q$ , whereas the curves $3$ and $4$ correspond to
the same quantities with $\gamma_c=3\gamma_q$. The line $5$ in
Fig.2 corresponds to the average square for the transverse momentum
of $K^+$ produced in the free  $p+p$ collisions
$<p_t^2>_{K^+}^{NN}=0.14  \ GeV/c^2$. 
As our calculations show, the temperature dependence for\\
$<p_{K^+,t}^2>_{hh}^{AA}$ is rather
weak in the interval $T=100 - 150$ MeV. 

\begin{figure}[t]
\rotatebox{270}%
{\epsfig{file=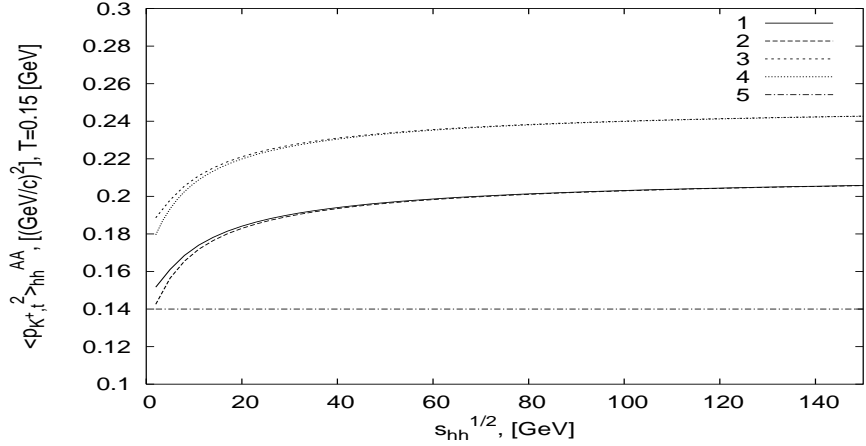,height=12.cm,width=6.cm }}
\caption[Fig.2]{Average square for the transverse momentum of $K^+$-meson produced from the
interaction of two hadrons one of them is in the
 equilibrated fireball
as a function of $\sqrt{s_{hh}}$ at $T=150$ MeV.}
\end{figure}

As is evident from Fig.2, the obtained results are sensitive to
the mass value of a hadron which is locally equilibrated with the
surrounding nuclear matter at $\sqrt{s_{hh}}\leq 10 (GeV)$.

We found that the quark distribution in a hadron depends on
the fireball temperature $T$. At any $T$ the average transverse momentum squared
of a quark grows and then saturates when $\sqrt{s_{hh}}$ increases. Numerically this saturation 
property depends on $T$. It leads to a similar energy dependence for the average transverse 
momentum squared of hadron $h_1$ $<p_{h_1,t}^2>_{hh}^{AA}$. 
The saturation property for $<p_{h_1,t}^2>_{hh}^{AA}$ depends also 
on  the temperature $T$ and it is very sensitive to the dynamics of hadronization.
As an example, we studied the energy dependence of the inverse slope of transverse mass spectrum
of $K$-mesons produced in central heavy-ion collisions and got its energy dependence
qualitatively similar observed to one  experimentally. We guess that our assumption on the thermodynamical 
equilibrium of hadrons given by eq.(\ref{def:hdmp}) can be applied for heavy nuclei only
and not for the early interaction stage.

\section*{Acknowledgments}

The authors are grateful for very useful discussions with
P.Braun-Munzinger, K.A.Bugaev, W.Cassing, A.V.Efremov, 
M.Gazdzicki, S.B.Gerasimov, M.I.Gorenstein, Yu.B.Ivanov,
A.B.Kaidalov and O.V.Teryaev. This work was supported in part by RFBR 
Grant N 05-02-17695 and by the special program of the Ministry 
of Education and Science of the Russian Federation (grant
RNP.2.1.1.5409).

\end{document}